\documentclass[%
 reprint,
 amsmath,amssymb,
 aps,
prl]{revtex4-1}

\usepackage{subfigure}
\usepackage{graphicx}
\usepackage{dcolumn}
\usepackage{bm}

\begin{document}

\preprint{APS/123-QED}

\title{Unconventional Spin Hall Effect and Axial Current Generation in a Dirac Semimetal}

\author{Nobuyuki Okuma}
\email{okuma@hosi.phys.s.u-tokyo.ac.jp}
\author{Masao Ogata}%
\affiliation{%
 Department of Physics, University of Tokyo, Hongo 7-3-1, Tokyo 113-0033, Japan
}%

%
\date{\today}

\begin{abstract} 
We investigate electrical transport in a three-dimensional massless Dirac fermion model that describes a Dirac semimetal state realized in topological materials.
We derive a set of interdependent diffusion equations with 8 local degrees of freedom, including the electric charge density and the spin density, that respond to an external electric field.
By solving the diffusion equations for a system with a boundary, we demonstrate that a spin Hall effect with spin accumulation occurs even though the conventional spin current operator is zero.
The Noether current associated with chiral symmetry, known as the axial current, is also discussed.
We demonstrate that the axial current flows near the boundary and that it is perpendicular to the electric current.

\begin{description}
\item[PACS numbers]
72.25.-b, 85.75.-d, 72.10.-d, 75.76.+j, 71.70.Ej
\end{description}

\end{abstract}

\maketitle


$Introduction$.\textemdash Massless Dirac fermions (MDFs) have been widely studied not only in particle physics but also in condensed matter physics.
In the latter context, MDF models describe materials whose conduction and valence bands touch with linear dispersion at isolated (Dirac) points in momentum space.
Such a band structure has been found in two-dimensional systems such as graphene \cite{castro} and topological insulator surface states \cite{hasan,xlq}.
In recent years, three-dimensional analogues of these materials called Dirac semimetals (DSMs) have been theoretically predicted \cite{singh,young,wang1,wang2}.
Na$_3$Bi \cite{liu,xu} and Cd$_3$As$_2$ \cite{neupane,borisenko} are thought to be experimentally realized DSMs with symmetry protected Dirac points.
The DSM state is also believed to be realized in topological materials such as TlBi(S$_{1-x}$Se$_x$)$_2$ \cite{sato,xu2,novak} and (Bi$_{1-x}$In$_x$)$_2$Se$_3$ \cite{brahlek,wu}.

One of the important differences between two- and three-dimensional MDF systems is the number of local degrees of freedom (DOFs) such as the electric charge density and the spin density.   
For a Dirac fermion field $\psi$, the low energy effective Hamiltonian in $d$ spatial dimensions is given by
\begin{align}
H=\int \frac{d^dp}{(2\pi)^d}\psi^{\dagger}_{\bm{p}}\hat{\mathcal{H}}_{\bm{p}}\psi_{\bm{p}}=\int \frac{d^dp}{(2\pi)^d}\psi^{\dagger}_{\bm{p}}\left[\sum_{i=1}^{d}v_ip_i\hat{\alpha}_i\right]\psi_{\bm{p}},\label{MDham}
\end{align}
where $\hat{\mathcal{H}}$ is a $2^{d-1}\times2^{d-1}$ Hermitian matrix, $\psi$ is a $2^{d-1}$-component spinor, 
$\bm{v}=(v_1,\cdots, v_d)$ is the Fermi velocity, $\bm{p}=(p_1,\cdots, p_d)$ is the crystal momentum measured from the Dirac point, $\hat{\alpha}_i$ are the alpha matrices obeying the Clifford algebra $\{\hat{\alpha}_{\mu},\hat{\alpha}_{\nu}\}=2\delta_{\mu\nu}$ ($\mu,\nu=0,\cdots,d$), and we use $\hbar=1$ henceforth.
Thus, in two dimensions, the largest number of linearly independent local operators \cite{physical}, which is the number of the independent components of a $2\times2$ Hermitian matrix, is 4, e.g., the particle density $N=\psi^{\dagger}\psi$ and the spin density $\bm{S}=\psi^{\dagger}(\hat{S}_x, \hat{S}_y, \hat{S}_z)\psi$ in a topological insulator surface state \cite{burkovhawthorn}.
In three dimensions, on the other hand, that number is 16.
It follows that the potential for interesting new effects is greater in three dimensions.

\begin{figure}[bp]
\begin{center}
　　　\includegraphics[width=7cm,angle=0,clip]{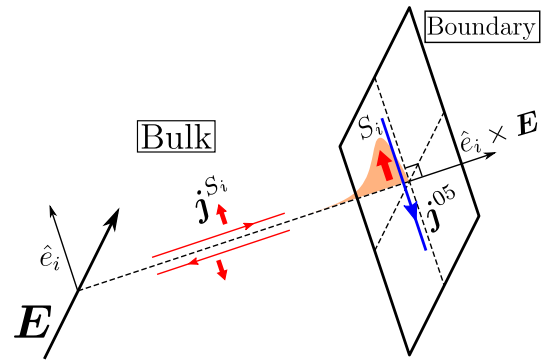}
　　　\caption{Schematic illustration of electrical transport in the DSM state realized in TlBi(S$_{1-x}$Se$_x$)$_2$. The electric field $\bm{E}$ induces the $i$-component spin current $\bm{j}^{S_i}$ in the $\hat{e}_i\times\bm{E}$ direction in the bulk, and the $i$-component spin density $S_i$ accumulates near the boundary with the normal vector $\hat{e}_i\times\bm{E}$. The axial current $\bm{j}^{05}$ flows near the boundary in the $-\hat{e}_i$ direction. See text for a detailed discussion.}
　　　\label{fig1}
\end{center}
\end{figure}

In this paper, we derive a set of interdependent diffusion equations in the presence of an electric field, involving 8 local DOFs for a DSM state in topological materials.
We show that an unconventional spin Hall effect occurs in the bulk of the system, while the axial current, which is unique to massless Dirac fermion systems, flows near the boundary (See Fig. $\ref{fig1}$).

$Spin$ $transport$ $in$ $a$ $Dirac$ $Semimetal$.\textemdash 
We start with a $d=3$ MDF Hamiltonian Eq. (\ref{MDham}) with $v_i=v$. We assume an explicit representation of the alpha matrices: $\hat{\alpha}_i=\hat{\sigma}_i\otimes\hat{\tau}_1$, with $\hat{\sigma}_i$ and $\hat{\tau}_i$ being the Pauli matrices in spin and orbital space, respectively. 
Although the Hamiltonian includes spin-orbit interaction terms, the conventional spin Hall coefficient $\sigma^{S_z}_{xy}$ is zero because the conventional spin current operator is given by
\begin{align}
\hat{j}^{S_z}_{y}=\frac{\left\{\hat{S}_z,\frac{\partial \hat{\mathcal{H}}}{\partial p_y}\right\}}{2}=0,
\end{align}
which apparently indicates the absence of non-trivial spin transport.
When spin is not conserved, however, the conventional spin current operator has no theoretical foundation and often leads to unphysical results.
For instance, Rashba \cite{rashba} constructed an example in which the conventional spin current is non-zero even in equilibrium.
The definition of the spin current operator is still controversial, and there are other proposals of the definition \cite{mnz,shi}.

Instead of spin current operators, we herein use the spin density, which is always a well-defined observable \cite{macdonald,mishchenko}.
To investigate the non-equilibrium dynamics of the spin density, we start with a quantum kinetic equation (QKE) \cite{mishchenko,rammer}, 
which allows us to perform a space-dependent analysis of local DOFs such as the spin density.  
We introduce a momentum- and energy-dependent density matrix $\hat{g}(\bm{p},\epsilon,\bm{x},t)$.
In the Born approximation for non-magnetic impurity scattering, the QKE for $\hat{g}(\bm{p},\epsilon,\bm{x},t)$ in a uniform electric field $\bm{E}$ is given by \cite{mishchenko}
\begin{align}
\partial_t\hat{g}+\frac{1}{2}\left\{\widetilde{\partial}_i\hat{g},\frac{\partial \hat{\mathcal{H}}}{\partial p_i}\right\}+i[\hat{\mathcal{H}},\hat{g}]  
=-\frac{1}{\tau}\hat{g}+\frac{i}{\tau}\left(\hat{G}^R\hat{\rho}-\hat{\rho} \hat{G}^A   \right),\label{qke}
\end{align}
where $\widetilde{\partial}_{i}=\partial/\partial r_i+eE_i\partial_{\epsilon}$, $\hat{G}^{R,A}=[\epsilon-(\hat{\mathcal{H}}-\mu)\pm\frac{i}{2\tau}]^{-1}$ are the retarded and advanced  Green's functions, $1/\tau$ is the impurity scattering rate, $\mu>0$ is the chemical potential, and
\begin{align}
\hat{\rho}({\epsilon},\bm{x},t)=\frac{1}{\pi\nu}\int\frac{d^3p}{(2\pi)^3}\hat{g}(\bm{p},\epsilon,\bm{x},t)
\end{align}
is the energy-dependent density matrix. Here, $\nu=\mu^2/2\pi^2v^3$ is the density of states per band at the Fermi energy.  

To obtain the diffusion equations, we solve Eq. (\ref{qke}) approximately.
For convenience, we introduce the time Fourier transforms $\hat{g}_{\omega}(\bm{p},\epsilon,\bm{x})$ and $\hat{\rho}_{\omega}(\epsilon,\bm{x})$.
The Fourier transform of Eq. (\ref{qke}) can be formally solved as 
\begin{align}
\hat{g}_{\omega}&=i\frac{(2E_{\bm{p}}^2-\Omega^2)\hat{F}+2E_{\bm{p}}^2\hat{\alpha}_{\theta\phi}\hat{F}\hat{\alpha}_{\theta\phi}-\Omega E_{\bm{p}}\left[\hat{\alpha}_{\theta\phi},\hat{F}\right]}{\Omega\left(4E_{\bm{p}}^2-\Omega^2\right)}\notag\\
&\equiv \hat{g}^{(0)}_{\omega}+\hat{G}_{\mathrm{grad}}[\hat{g}_{\omega}],\label{iteration}
\end{align}
where $\hat{F}=\hat{F}_0+\hat{F}_{\mathrm{grad}}$, $\hat{F}_0=i/\tau(\hat{G}^R\hat{\rho}_{\omega}-\hat{\rho}_{\omega}\hat{G}^A)$, $\hat{F}_{\mathrm{grad}}=-\{\widetilde{\partial}_i\hat{g}_{\omega}, \partial \hat{\mathcal{H}}/\partial p_i  \}/2$, $\hat{\alpha}_{\theta\phi}=\mathrm{sin}\theta\mathrm{cos}\phi\hat{\alpha}_1+\mathrm{sin}\theta\mathrm{sin}\phi\hat{\alpha}_2+\mathrm{cos}\theta\hat{\alpha}_3$, $\Omega=\omega+i/\tau$, and $E_{\bm{p}}=v|\bm{p}|$. $\hat{g}^{(0)}_{\omega}$ and $\hat{G}_{\mathrm{grad}}[\hat{g}_{\omega}]$ are $\hat{F}_0$- and $\hat{F}_{\mathrm{grad}}$- dependent parts of the first line,  respectively. $\theta$ and $\phi$ are the polar and azimuthal angles of the momentum $\bm{p}$.
Assuming $\widetilde{\partial}_i \ll p_F\equiv \mu/v$, we regard $\hat{G}_{\mathrm{grad}}$ as a perturbation and perform a gradient expansion \cite{mishchenko}.
Solving Eq. (\ref{iteration}) with respect to $\hat{g}_{\omega}$ by a second order iteration, integrating over both $\epsilon$ and $\bm{p}$, and performing an inverse Fourier transform with respect to $\omega$ \cite{relax}, we obtain the diffusion equation for the density matrix $\hat{\mathcal{D}}(\bm{x},t)\equiv \nu\int d\epsilon\hat{\rho}(\epsilon,\bm{x},t)$ \cite{mishchenko}.

For convenience, we decompose the density matrix $\hat{\mathcal{D}}$ into 16 linearly independent components:
\begin{align}
\hat{\mathcal{D}}=&\frac{1}{4}\left(N\hat{1}+\sum_{a=0,1,2,3,5}\rho^a\hat{\alpha}_a+\sum_{\substack{a<b,\\a,b=0,1,2,3,5 } }\rho^{ab}(i\hat{\alpha}_a\hat{\alpha}_b)\right).
\end{align}
Here, $\hat{\alpha}_5=\hat{\alpha}_0\hat{\alpha}_1\hat{\alpha}_2\hat{\alpha}_3$, and we define 16 local DOFs: $N,\ \rho^a,$ and $\rho^{ab}$.
In the DSM state with $\hat{\alpha}_i=\hat{\sigma}_i\otimes\hat{\tau}_1$, the spin operator is $\hat{\bm{S}}\equiv (\hat{\bm{\sigma}}\otimes\hat{1})/2=(-i\hat{\alpha}_2\hat{\alpha}_3,i\hat{\alpha}_1\hat{\alpha}_3,-i\hat{\alpha}_1\hat{\alpha}_2)/2$, and the spin density is $\bm{S}(\bm{x},t)=(-\rho^{23}(\bm{x},t),\rho^{13}(\bm{x},t),-\rho^{12}(\bm{x},t))/2$. By using these local DOFs, we obtain a set of interdependent diffusion equations with 16 local DOFs. In practice, however, we can limit the discussion to the following closed equations for 8 local DOFs including the particle density $N(\bm{x},t)$ and the spin density $\bm{S}(\bm{x},t)$:
\begin{subequations}
\begin{align}
\frac{\partial N}{\partial t}=&-\nabla\cdot\left[-D\nabla N+De(2\nu) \bm{E}\right]-\frac{v}{3}\nabla\cdot\bm{\rho}, \label{a}\\
\frac{\partial \rho^{05}}{\partial t}=&D\nabla^2 \rho^{05}+\frac{2v}{3}\nabla\cdot\bm{S},\label{b}\\
\frac{\partial \bm{\rho}}{\partial t}=&\frac{D}{5}\nabla^2 \bm{\rho}+\frac{2D}{5}\nabla\left(\nabla\cdot\bm{\rho}\right)-\frac{v}{3}\nabla N+\frac{ve(2\nu)}{3}\bm{E}\notag\\
&+\frac{v}{3\mu\tau}\nabla\times\bm{S}-\frac{\bm{\rho}}{\left(\frac{3\tau}{2}\right)},\label{c}\\
\frac{\partial \bm{\bm{S}}}{\partial t}=&\frac{D}{5}\nabla^2 \bm{\bm{S}}+\frac{2D}{5}\nabla\left(\nabla\cdot\bm{\bm{S}}\right)+\frac{v}{6}\nabla \rho^{05}\notag\\
&+\frac{v}{12\mu\tau}\nabla\times\bm{\rho}-\frac{\bm{S}}{\left(\frac{3\tau}{2}\right)},\label{d}
\end{align}\label{mainresult}
\end{subequations}
where $D=v^2\tau/3$ is the diffusion constant, and $\bm{\rho}=(\rho^1,\rho^2,\rho^3)$.
Here, we have derived Eqs. (\ref{mainresult}) in the quasi particle approximation ($1/\tau\ll\mu$) and have used only the zeroth and first order terms of the electric field. 
As a result, the electric field only appears in the form $\nabla N-e(2\nu)\bm{E}$.
Note that $\rho_0$, $\rho_5$, $\rho_{0i}$, and $\rho_{i5}$ do not respond to the first-order electric field. 
Thus, we ignore these local DOFs henceforth.

Equations (\ref{a}) and (\ref{b}) have the form of a continuity equation $\partial_t j_0^{N}=-\nabla\cdot\bm{j}^{N}$, where $j^N_{\mu}$ is the Noether four-current. 
These relations originate from the fact that MDF systems have U(1) gauge symmetry, and also chiral symmetry, as will be discussed later. 
From Eq. (\ref{a}), the electric current $\bm{j}$, which is the Noether current associated with U(1) gauge symmetry, can be written as
\begin{align}
\bm{j}=-D\nabla N+De(2\nu)\bm{E}+\bm{j}_a,\label{elcurrent}
\end{align}
where $\bm{j}_a\equiv v\bm{\rho}/3$, and we normalize the electric current by the elementary charge $e$ henceforth.
The first and second terms are the diffusion current and the usual drift current, respectively. The third term is an additional current that is absent in the electron gas model with quadratic dispersion.
We can interpret $\bm{j}_a$ as an impurity vertex correction to the longitudinal current in the Kubo formalism \cite{vertex}.
The existence of the vertex correction term is a consequence of the particle conservation law, which holds in our formalism.

To investigate the spin Hall effect, we consider a steady state ($\partial/\partial t=0$) under physical boundary conditions.
The solution of the diffusion equations depends on the choice of boundary conditions \cite{tse,galitski}. 
We here assume that every local DOF is zero on boundaries.
Under the boundary conditions, we have the following relations in the steady state:
\begin{align}
&N=0,\ \rho^{05}=0,\ \nabla\cdot\bm{\rho}=0,\ \nabla\cdot\bm{S}=0,\notag\\
&0=\frac{\partial \rho_i}{\partial t}=\frac{D}{5}\nabla^2\rho_i+\frac{ve(2\nu)}{3}E_i+\frac{v}{3\mu\tau}\nabla\cdot(\bm{S}\times\hat{e}_i)-\frac{\rho_i}{\left(\frac{3\tau}{2}\right)},\notag\\
&0=\frac{\partial S_i}{\partial t}=\frac{D}{5}\nabla^2S_i+\frac{v}{12\mu\tau}\nabla\cdot(\bm{\rho}\times\hat{e}_i)-\frac{S_i}{\left(\frac{3\tau}{2}\right)},\label{spindiff}
\end{align}
where $S_i$ and $\rho_i$ are scalar projections of $\bm{S}$ and $\bm{\rho}$ on a unit vector $\hat{e}_i$, respectively.
The phenomenological spin diffusion equation for $S_i$ is given by 
\begin{align}
\frac{\partial S_i}{\partial t}=D_s\nabla^2S_i-\nabla\cdot\bm{j}^{S_i}-\frac{S_i}{\tau_s},\label{pheno}
\end{align}
where $D_s$ is the spin diffusion constant, $\tau_s$ is the spin relaxation time, and $\bm{j}^{S_i}$ is the $i$-component spin current. Comparing Eqs. (\ref{spindiff}) with Eq. (\ref{pheno}), we obtain the following expressions:
\begin{align}
\bm{j}^{S_i}=\frac{1}{4\mu\tau}\hat{e}_i\times\bm{j}_a,\ D_s=\frac{D}{5},\ \tau_s=\frac{3\tau}{2}.\label{explicitform}
\end{align}
Note that we do not use any definition of the spin current operator to determine the spin current expression.
From Eqs. (\ref{explicitform}), the spin current $\bm{j}^{S_i}$ is closely related to the additional current $\bm{j}_a$.
In the bulk ($\nabla=0$), we obtain non-zero polarization of $\bm{\rho}$ from Eqs. (\ref{spindiff}):
\begin{align}
\bm{\rho}(\mathrm{bulk})=\tau v e \nu \bm{E},\label{rhovalue}
\end{align}
which leads to the non-zero additional current $\bm{j}_a=v\bm{\rho}/3$. Thus, the $z$-component spin Hall coefficient $\sigma^{S_z}_{xy}$ is given by
\begin{align}
\sigma^{S_z}_{xy}\equiv \frac{j^{S_z}_y(\mathrm{bulk})}{E_x}=\frac{e}{24\pi^2}p_F.\label{shc}
\end{align}
It is interesting to note that the $z$-component spin Hall coefficient is non-zero even though the conventional spin current operator $\hat{j}^{S_z}_y$ is zero.
The origin of this spin current is $[\hat{\alpha}_{\theta\phi},\hat{F}]$ in Eq. (\ref{iteration}), which describes the $\tau$-independent correction to the $\tau$-dependent transport.
Our result is an example of an unconventional spin Hall effect that can not be predicted by the Kubo formula for the conventional spin current operator.

\begin{figure}[tbp]
\begin{center}
　　　\includegraphics[width=7cm,angle=0,clip]{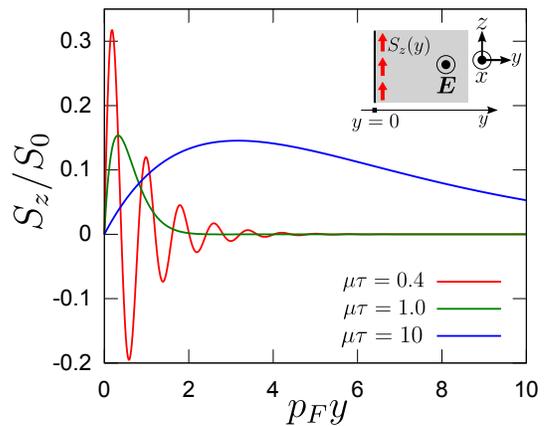}
　　　\caption{The $z$-component spin density normalized by $S_0=ve\nu E_x/2\mu$ as a function of $p_Fy$ for various $\mu\tau$, calculated using the solution given by Eqs. ($\ref{spinaccum}$).}
　　　\label{fig2}
\end{center}
\end{figure}

Since this spin Hall effect is a diffusive phenomenon, the spin current causes spin accumulation near the boundary, as shown below. 
We now solve the diffusion equations ($\ref{spindiff}$) in the presence of an electric field $\bm{E}=(E_x,0,0)$ for $y\geq0$. 
The physical solution satisfying the boundary conditions $\bm{\rho}(y=0)=\bm{S}(y=0)=0$ is given by 
\begin{align}
\rho_x&=\tau v e \nu E_x\left[1-\mathrm{exp}\left(-\frac{y}{l_s}\right)\mathrm{cos}\left(\frac{y}{l_{osc}}\right)\right],\ \rho_y=\rho_z=0,\notag\\
S_z&=\frac{\tau v e \nu E_x}{2}\mathrm{exp}\left(-\frac{y}{l_s}\right)\mathrm{sin}\left(\frac{y}{l_{osc}}\right),\ S_x=S_y=0,\label{spinaccum}
\end{align}
where $l_s=v\tau/\sqrt{10-25/(16\mu^2\tau^2)}$ is the spin diffusion length, and $l_{osc}=4\mu\tau v\tau/5$ is the oscillation length of local DOFs.
The $z$-component spin density distribution is plotted for various $\mu\tau$ in Fig. $\ref{fig2}$.
Although the solution given by Eqs. ($\ref{spinaccum}$) has oscillations, cancellation of the net spin accumulation is negligibly small for sufficiently large $\mu\tau$, where the quasi particle approximation ($\mu\tau\gg1$) is valid.
Note that the accumulated spin is perpendicular to the electric field and parallel to the boundary, while that in the Rashba model, which is a typical model for the spin Hall effect, is not \cite{galitski}. 

For a qualitative estimate, we use the following typical values for the DSM state realized in TlBi(S$_{1-x}$Se$_x$)$_2$ \cite{xu2,novak}: the scattering time $\tau\sim10^{-13}$ s, the Fermi velocity $v\sim10^5$ m/s, and the Fermi wavenumber $p_F\sim10^9$ $\mathrm{m}^{-1}$. 
In this material, the quasi particle approximation is justified since $\mu\tau/\hbar\sim10^1>1$.
By using Eq. (\ref{shc}), we obtain the spin Hall coefficient $\sigma_{xy}^{S_z}\sim 10^1 (\hbar/\mathrm{e})(\Omega\ \mathrm{cm})^{-1}$, which is an order of magnitude larger than the typical value for semiconductors \cite{matsuzaka}. 
We also obtain the spin diffusion length $l_s\sim 10^1$ nm. Thus, our unconventional spin Hall effect is expected to be observed, as in standard spin Hall materials.

Another interesting feature of DSMs is transport related to chiral symmetry \cite{burkovaxial,taguchi}.
Three-dimensional MDF systems are invariant under a chiral transformation $\psi\rightarrow \mathrm{e}^{i\theta\hat{\gamma}_5}\psi$, where $\hat{\gamma}_5\equiv i\hat{\alpha}_0\hat{\alpha}_5$. 
From Eq. (\ref{b}), the Noether current associated with this symmetry, known as the axial current $\bm{j}^{05}$ in quantum field theory, is given by
\begin{align}
\bm{j}^{05}=-\nabla\rho^{05}-\frac{2v}{3}\bm{S},
\end{align}
where $\rho^{05}$ is the axial charge density.
In the steady state described by Eqs. (\ref{spindiff}), $\bm{S}$ is a divergenceless vector field, the bulk spin density $\bm{S}(\mathrm{bulk})=0$, and $\bm{j}^{05}$ is proportional to $\bm{S}$ since $\rho^{05}=0$.
Thus, we obtain the following expression:
\begin{align}
&\bm{j}^{05}=\nabla\times\bm{B}^{05},
\end{align}
where $\bm{B}^{05}$ is equivalent to the ``magnetic field" for $\rho^{05}$. Note that the axial current has a similar form to the persistent electric current in the presence of a real magnetic field.
The axial current in the bulk is zero, while that near the boundary is non-zero. 
In terms of the chiral symmetry of the three-dimensional MDF system, the spin accumulation can be interpreted as the axial current generation near the boundary.
The direction of the axial current is opposite to the accumulated spin, which is perpendicular to the electric field and parallel to the boundary, as discussed above (See Fig. $\ref{fig1}$).

$Generalization\ to\ other\ Dirac\ Semimetals$.\textemdash 
Because the derivation of the diffusion equations relies only on the Clifford algebra $\{\hat{\alpha}_{\mu},\hat{\alpha}_{\nu}\}=2\delta_{\mu\nu}$, which is a general relationship of three-dimensional MDF systems, the above discussion for the specific DSM can be straightforwardly generalized to other DSMs.
It is important to note, however, that the physical meaning of $\bm{\rho}$, $\bm{S}$, and $\rho^{05}$ depends on the type of DSM.
For instance, $\bm{S}=(-\rho^{23},\rho^{13},-\rho^{12})/2$ is not always the real spin density, whereas Eqs. (\ref{mainresult}) hold for any representation of the alpha matrices.
When the axial charge $\rho^{05}$ is an experimental observable such as the spin density, the axial current can be detected.

Finally, we discuss the recent DSM candidate $\mathrm{Cd}_3\mathrm{As}_2$ \cite{wang2,neupane,borisenko}.
In the effective Hamiltonian around one Dirac point (valley), the alpha matrices have the following representations \cite{wang2}: $\hat{\alpha}_1=\hat{\sigma}_3\otimes\hat{\tau}_1$, $\hat{\alpha}_2=-\hat{1}\otimes\hat{\tau}_2$, and $\hat{\alpha}_3=\hat{1}\otimes\hat{\tau}_3$, with $\hat{\sigma}_i$ and $\hat{\tau}_i$ being the Pauli
matrices in spin and orbital space, respectively. In this representation, the spin Hall effect no longer occurs since $\bm{S}$ is not the spin density.
Instead, the axial charge is proportional to the $z$-component spin density since $\hat{\gamma}_5=\hat{\sigma}_3\otimes\hat{1}$.
Thus, the axial current is the conserved $z$-component spin current in this effective model.
Because the conserved spin current is exactly cancelled by a spin current in the opposite direction from the other Dirac point (valley) in realistic materials, it is necessary to create a chemical potential difference between the two valleys in order to detect the spin current.

$Summary$.\textemdash 
We have derived a set of diffusion equations with 8 local degrees of freedom in a three-dimensional massless Dirac fermion model that describes the Dirac semimetal state realized in $\mathrm{Tl}\mathrm{Bi}(\mathrm{S}_{1-x}\mathrm{Se}_x)_2$. We have found that an unconventional spin Hall effect in which the conventional spin current operator is zero occurs in the bulk, while the axial current flows near the boundary.
Because the derivation of the diffusion equations relies only on general properties of three-dimensional massless Dirac fermion models, our discussions can be straightforwardly generalized to other Dirac semimetals such as the recent Dirac semimetal candidate $\mathrm{Cd}_3\mathrm{As}_2$.

We acknowledge many fruitful discussions with Allan H. MacDonald, Hideo Aoki, Gen Tatara, Yusuke Horinouchi, Tomonari Mizoguchi, and Joel Foo.
This work was supported by Grant-in-Aid for Scientific Research (A) (No. 15H02108) from Japan Society for the Promotion of Science.
N. O. was supported by the Japan Society for the Promotion of Science through Program for Leading Graduate Schools (MERIT).


\begin{thebibliography}{9}
\bibitem{physical}
Here, we consider operators that act only on spinor indices.
\bibitem{relax}Although the final form of $\hat{g}_{\omega}$ has a complicated $\omega-$dependence, we can perform the inverse Fourier transform in a quasistationary regime ($\omega\tau\ll1$), as discussed in Ref. \cite{mishchenko}.
\bibitem{vertex} The vertex correction to the longitudinal current is given by $\delta\bm{j}=De\nu\bm{E}$ in the Kubo formalism.
This is exactly the same as the bulk additional current $\bm{j}_a$ derived from Eq. (\ref{rhovalue}).
\end{thebibliography}
\end{document}